# Surface nanoscale axial photonics: Robust fabrication of high quality factor microresonators

M. Sumetsky[*], D. J. DiGiovanni, Y. Dulashko, J. M. Fini, X. Liu, E. M. Monberg, and T. F. Taunay

*OFS Laboratories, 19 Schoolhouse Road, Somerset, NJ 08873*
*\*Corresponding author: M. Sumetsky sumetski@ofsoptics.com*

Recently introduced Surface Nanoscale Axial Photonics (SNAP) makes it possible to fabricate high Q-factor microresonators and other photonic microdevices by dramatically small deformation of the optical fiber surface. To become a practical and robust technology, the SNAP platform requires methods enabling reproducible modification of the optical fiber radius at nanoscale. In this paper, we demonstrate super-accurate fabrication of high Q-factor microresonators by nanoscale modification of the optical fiber radius and refractive index using the $CO_2$ laser and the UV excimer laser beam exposures. The achieved fabrication accuracy is better than 2 angstroms in variation of the effective fiber radius.

Surface nanoscale axial photonics (SNAP) is concerned with microscopic optical devices created by smooth and dramatically small *nanoscale* variation of the optical fiber radius and/or equivalent variation of its refractive index [1,2]. SNAP is based on whispering gallery modes (WGMs), which slowly propagate along the fiber axis and circulate around its surface. Unlike slow light in photonic crystals, which is introduced with the periodic modulation of the refractive index [3], the slow axial propagation of light in SNAP is insured automatically by periodic revolution around the fiber surface. The direction of light propagation considered in SNAP is primarily azimuthal so that the axial propagation is naturally slow.

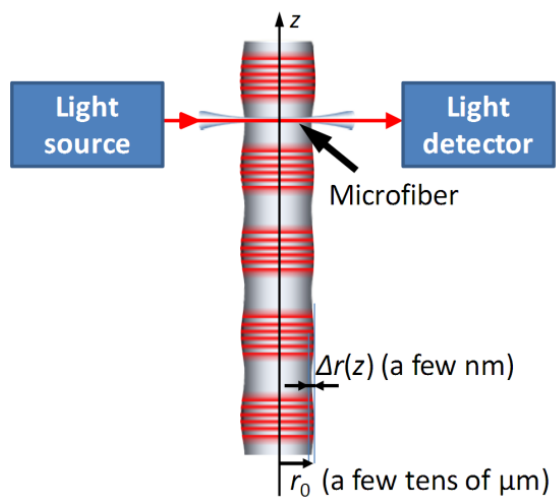

Fig. 1. (Color online) Illustration of a SNAP device composed of a sequence of bottle microresonators formed by nanoscale variation of the optical fiber radius. Light is coupled into the microresonators with a microfiber, which is connected to the light source and detector.

There is both fundamental and applied interest in the development of SNAP. The SNAP devices are fabricated of drawn silica and, hence, exhibit very small losses and high Q-factors, similar to those of silica WGM microresonators [4]. In addition, the characteristic axial wavelength of these devices is much greater than the wavelength of light, which makes them convenient for investigation of fundamental properties of light, e.g., tunneling, halting by a point source, and formation of dark states [1,2]. For the same reason, a large axial wavelength simplifies accurate fabrication of high Q-factor microresonators and other microdevices at the surface of a fiber. Typically, the SNAP photonic elements have tens of μm dimensions and a record small attenuation of light with a Q-factor in excess of $10^6$ [1,2]. This suggests SNAP as a potential platform for miniature integrated photonic circuits having attenuation of light which is orders of magnitudes smaller than that in microscopic devices fabricated with the existing photonic platforms [5-7].

This paper solves a challenging problem of accurate and reproducible modification of the optical fiber effective radius at nanoscale, which is critical for establishing of SNAP as a practical technology. We propose and demonstrate the fabrication of high Q-factor SNAP microresonators (Fig. 1) based on IR ($CO_2$ laser) and UV (248 nm excimer laser) beam exposures.

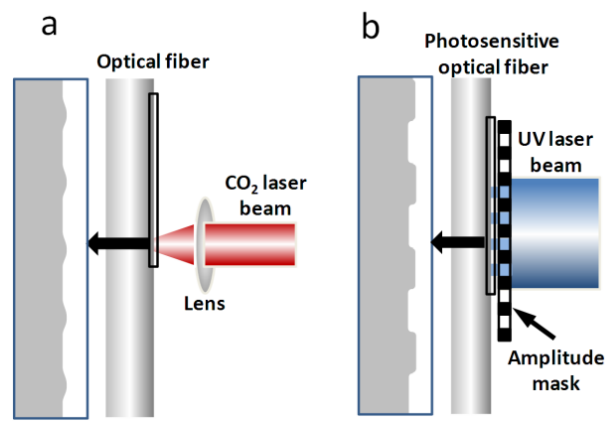

Fig. 2. (Color online) Illustration of the setups for nanoscale modification of the optical fiber radius and refractive index: (a) – with a focused $CO_2$ laser beam exposure; (b) – with a UV laser beam exposure through an amplitude mask.

In our experiments illustrated in Fig. 2, the local perturbations of the optical fiber shape and/or refractive index introduced by the IR and UV radiation are axially asymmetric due to the asymmetric position of the laser beam with respect to the fiber and attenuation of radiation in silica. Usually, these small and smooth perturbations result in negligible deformation of the WGM distribution along the transverse cross-section of the fiber. Then, in cylindrical coordinates $(z,\rho,\varphi)$, the field components with angular momentum $m$ retain the separable form $\Psi(z,\lambda)A(\rho)\exp(im\varphi)$. However, these perturbations may strongly affect the WGM amplitude variation, $\Psi(z,\lambda)$, and the WGM propagation constant, $\beta(z,\lambda)$, near a resonant wavelength $\lambda=\lambda_{res}$ [1,2]. The introduced asymmetric variation of the fiber radius, $r(\varphi,z)=r_0+\Delta r(\varphi,z)$, and its refractive index, $n(z,\rho,\varphi)=n_0+\Delta n(z,\rho,\varphi)$, enter the expression for the propagation constant only through their integral linear combination determined by the effective fiber radius, $\Delta r_{eff}(z)$:

$$\beta(z,\lambda)=\frac{2\pi n_0}{\lambda_{res}}\left(\frac{\Delta r_{eff}(z)}{r_0}-\frac{\lambda-\lambda_{res}}{\lambda_{res}}\right)^{1/2}. \quad (1)$$

where

$$\Delta r_{eff}(z)=\overline{\Delta r}(z)+\overline{\Delta n}(z)r_0/n_0, \quad (2)$$

and the average radius variation, $\overline{\Delta r}(z)$, and index variation, $\overline{\Delta n}(z)$, are:

$$\overline{\Delta r}(z)=\frac{1}{2\pi}\int_0^{2\pi}d\varphi\Delta r(z,\varphi),$$
$$\overline{\Delta n}(z)=\frac{1}{2\pi}\int_0^{2\pi}\int_0^{r_0}d\varphi d\rho\Delta n(z,\rho,\varphi)|A(\rho)|^2. \quad (3)$$

In Eq. (3), the WGM dependence on coordinate $\rho$ is normalized, $\int_0^\infty d\rho|A(\rho)|^2=1$. Eqs. (1)-(3) generalize the expression for the propagation constant $\beta(z,\lambda)$ obtained previously for the axially symmetric perturbations [1,2]. Similar to [1,2], the amplitude $\Psi(z,\lambda)$ of a WGM excited by a transverse microfiber (MF) positioned at $z=z_1$ (Fig. 1) is defined by the Schrödinger equation with the δ-function source:

$$A_{zz}+\beta^2(z,\lambda)A=C\delta(z-z_1), \quad (4)$$

where $C$ is the MF/SNAP fiber coupling parameter.

In our first experiment, small local variations of the density and refractive index of the optical fiber was introduced by annealing [8-11] performed with a $CO_2$ laser beam (Fig. 2(a)). The pure silica coreless fiber fabricated for this experiment had radius of 19 μm. Fig. 3(a) and (b) show series of extremely shallow bottle microresonators [12], which were created with focused laser beams having similar switching time (0.1 s) and exposure time (5 sec). The characterization of microresonators in Fig. 3 was performed with the method of scanning MF [13, 14, 2]. A transverse micron-diameter MF connected to the light source and detector (Fig. 1) was translated along the test fiber touching it in 20 μm steps where the resonant WGM transmission power spectra (plots in Fig. 3 shrunken along the vertical axis) were measured with a 3 pm wavelength resolution. These spectra allowed us to determine the variation of the effective fiber radius, $\Delta r_{eff}(z)$ determined by Eq. (2) (bold curves in Fig. 3). The measured spectra exhibited high Q-factor resonances with Q > $10^6$ (beyond the measurement resolution) for the MF positions close to the nodes of localized WGMs and in the evanescent regions. For other positions the resonances are wider due to the greater coupling to the MF. The series of 5 adjacent bottle microresonators shown in Fig. 3(a) were fabricated by successive exposures with 300 μm spacing. The relative beam powers of these exposures were set to 1, 0.9, 0.8, 0.7, and 0.6. The height of the first two microresonators is similar (23 nm) which indicates saturation of the refractive index and density variation for the given switching times and radiation power. Decreasing of the beam power in creation of the third, forth, and fifth microresonators led to reduction of their height to 20 nm, 12 nm, and 3 nm, respectively. In order to verify the accuracy and reproducibility of our technique, five adjacent bottle microresonators shown in Fig. 3(b) were fabricated with a similar laser beam power. The height of all five microresonators is close to 14 nm. From Fig. 3(b), the effective radius variation of these bottle resonators $R_{eff}$ can be approximated by the quadratic dependence, $\Delta r_{eff}(z) = z^2/(2R_{eff})$ with gigantic $R_{eff} = 0.24$ m. The axial free spectral range (FSR) of resonances in these microresonators calculated from Eq. (1), (2), and (4) is

$$\Delta \lambda_{FSR} = \lambda_{res}^2 (2\pi n_0)^{-1} (r_0 R_{eff})^{-1/2}. \qquad (4)$$

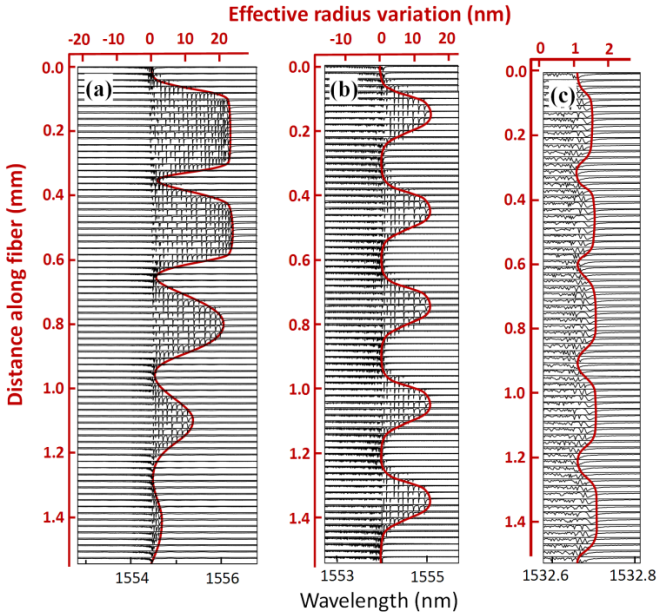

Fig. 3. (Color online) Series of microresonators created by annealing with (a) – different and (b) – similar $CO_2$ laser beam powers. (c) – Series of microresonators created with the UV beam exposure through an amplitude mask.

For $\lambda = 1.55$ μm, $n_0 = 1.46$, and $r_0 = 19$ μm this equation yields $\Delta \lambda_{FSR} = 0.123$ nm in good agreement with the experimental FSR found from Fig. 3(b), $\Delta \lambda_{FSR} = 0.12$ nm. Comparison of positions of resonances in different microresonators demonstrates remarkable reproducibility within 2 angstrom in the effective radius variation and 0.02 nm in wavelength variation.

In the second experiment, we modulated the effective radius variation of a photosensitive fiber with UV radiation, which affects both the refractive index and the density of the fiber (see, e.g., [15]). The Ge-doped coreless fiber fabricated for this experiment had 12.5% mol $GeO_2$ concentration and 14.5 μm radius. First, the fiber was exposed to 248 nm pulsed excimer laser beam through the amplitude mask with 300 μm period and 50% duty cycle. Characterization of the created series of ultra-shallow (~0.5 nm in effective radius variation) bottle microresonators with two resonances is shown in Fig. 3(c).

The introduced refractive index and density variation was axially asymmetric due to the substantial attenuation of the UV light in the doped fiber [16] with a maximum at the front fiber surface, and relatively small value at the back surface. The variation of the resonance positions and their widths observed along the length of each microresonator is explained by coupling to the MF which grows with the local amplitude of the resonant state. In order to increase the UV introduced effective radius variation further, the same Ge-doped fiber was hydrogen loaded [17]. A section of this fiber was exposed to the UV radiation through a 100 μm period 50% duty cycle phase mask followed by annealing. Fig. 4(a) and Fig. 4(b), (c) compare the effective radius variation of the adjacent unexposed and UV exposed sections. The effective fiber radius variation in Fig. 4(b), (c) is an order of magnitude greater than in the case of unloaded fiber, Fig. 3(c). The similar slopes in both figures are attributed to the original variation of the fiber radius. We suggest that this variation can be corrected with an additional UV and/or $CO_2$ laser exposures. After subtraction of the original fiber radius variation, we find the excellent reproducibility of the fabricated microresonators. In fact, the relative deviation of positions of their resonances is less than an angstrom in the effective radius variation and less than 0.01 nm in wavelength variation.

The axial dimension of the SNAP microresonators fabricated with the $CO_2$ beam exposure (Fig. 3(a) and (b)) $\Delta z_{loc} \sim 100$ μm, can be reduced by decreasing the focal size of the lens in Fig. 2(a) and variation of the beam power and exposure times. This dimension can be made as small as $\Delta z_{loc} \sim 10$ μm, which corresponds to the characteristic axial FSR $\Delta\lambda_{FSR}$ ~1 nm [18]. From Eq. (4), increasing of the FSR can be also achieved by increasing the height of microresonators keeping their width unchanged. Using the effect of saturation (see first microresonator in Fig. 3(a)) the shape of microresonators can be made close to rectangular. As opposed to the $CO_2$ laser heating method, the height of microresonators fabricated with the UV beam exposure is limited by the fiber photosensitivity. In this case, the axial dimensions and the shape of microresonators can be controlled by variation of the characteristics of the amplitude mask and the size of the beam.

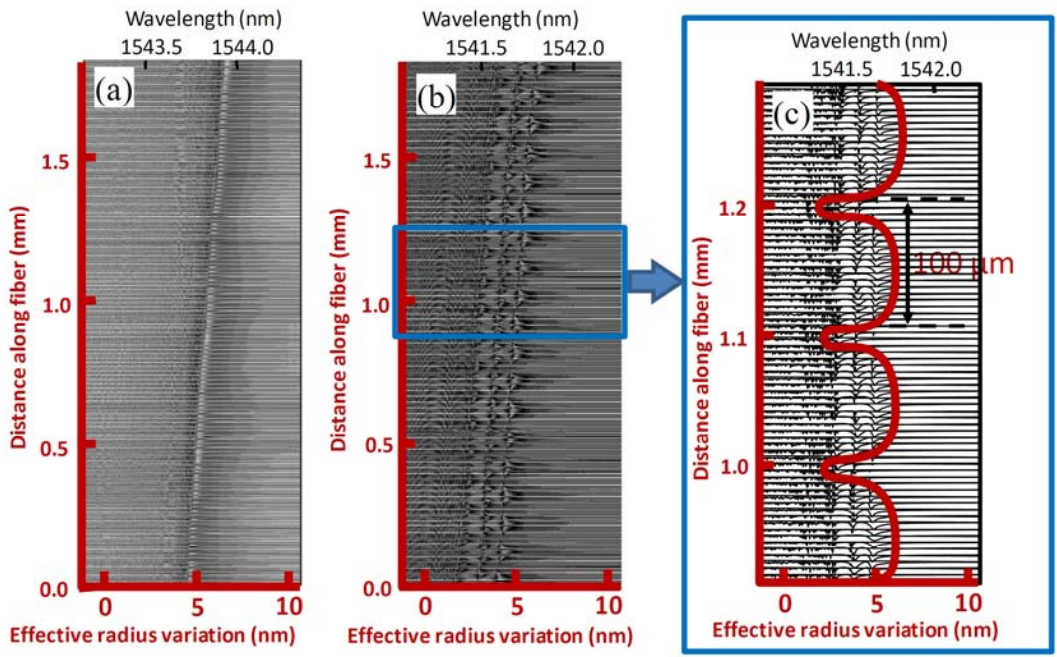

Fig. 4. (a) – Effective radius variation of a section of an unexposed fiber. (b) – Effective radius variation of a section of the same fiber adjacent to the section characterized in Fig. 4(a), which was exposed to the UV radiation through the 100 μm period 50% duty cycle phase mask. (c) – Magnified section of Fig. 4(b).

In summary, we have demonstrated the remarkably accurate, controllable, and reproducible fabrication of high Q-factor microresonators by nanoscale modification of the optical fiber effective radius. Two methods of nanoscale variation of the effective radius variation were developed. The first method, which is based on modification of the density and refractive index of the fiber material by annealing, is applicable to a variety of glass fibers. The second method, which is based on modification of the refractive index and density of the fiber by UV exposure, is applicable to photosensitive fibers. These results establish the robust technological foundation for the future development of the surface nanoscale axial photonics.

The authors are grateful to M. Fishteyn, J. W. Fleming, J. Porque, and P. S. Westbrook for useful consultations.